# Content-Based Personalized Recommender System Using Entity Embeddings

## Xavier Thomas

Manipal Institute of Technology, Udupi, Karnataka, India
xavier.thomas1@learner.manipal.edu

## Abstract

Recommender systems are a class of machine learning algorithms that provide relevant recommendations to a user based on the user's interaction with similar items or based on the content of the item. In settings where the content of the item is to be preserved, a content-based approach would be beneficial. This paper aims to highlight the advantages of the content-based approach through learned embeddings and leveraging these advantages to provide better and personalized movie recommendations based on user preferences to various movie features such as genre and keyword tags.

## Introduction

There are mainly two approaches to building a recommender system – Collaborative Filtering (CF) and Content-based (CB) recommending. CF systems work by collecting user feedback in the form of ratings for items in a given domain and exploit similarities and differences among profiles of several users in determining how to recommend an item. On the other hand, content-based methods provide recommendations by comparing representations of the content contained in an item to representations of content that interests the user. (Melville, Mooney and Nagarajan 2002)

The collaborative method creates a N x M matrix, with N users and M items. Wherein each user provides a rating for an item and this rating is stored as the element in the matrix. This matrix is utilized to provide similarity scores between users and recommends items that are highly rated by similar users. In the context of a movie recommendation system, which the paper describes. The collaborative method can lead to data sparsity, as a majority of the movies are unrated by the user. This method also undergoes a cold-start issue which causes newly added movies to have less significance in recommendations due to the lower number of ratings.

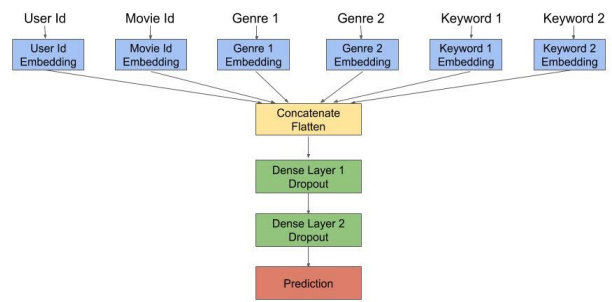

Figure 1: The proposed model takes the movie features as input, converts it into an embedding, and learns to predict the rating.

A content-based method does not suffer the problems stated above and can be used to build personalized recommender systems. This paper uses the MovieLens dataset to build this system.

## Methodology

The main idea behind building a content-based recommender system is to recommend similar items based on the content of the item. This paper uses the features genre and keywords attributed to each movie from the MovieLens Dataset. Today's Recommendation systems highly rely on similarity-based recommendations between users, which could fail to adapt to a user's unique taste. Learning an embedding space help to differentiate movies based on their content. This approach is utilized in finding movie and user embeddings so that similar movies and users are grouped respectively. Furthermore, this method is aimed to make recommender systems personalized by capturing the user's unique taste in the embedding space.

| User Id | 14 | 289 | 2 |
|---|---|---|---|
| Movie Id | 570 | 112 | 1127 |
| Genres | Drama, Crime | Horror, Crime | Drama, Romance |
| Genre 1 | 5 | 12 | 5 |
| Genre 2 | 11 | 11 | 21 |
| Keywords | Mafia, 1970s | Supernatural, Scary | Based on book, Love |
| Keyword 1 | 22 | 27 | 25 |
| Keyword 2 | 11 | 44 | 51 |
| Rating | 4.5 | 5 | 2.5 |

Table 1: Categorical variables Genre and Keywords are mapped to respective integer values.

**The proposed model** takes as inputs the User Id, Movie Id, Genre 1, Genre 2, Keyword 1, and Keyword 2. Where Genres and Keywords of the movie are categorical variables that are mapped to integer values as shown in the example in Table 1. The model learns to predict the Rating by converting the inputs into Embeddings and uses dense neural network architectures to output a predicted Rating.

In a more concise view, the model aims to learn a function

$$y = f(x_1, x_2, ...., x_n) \quad (1)$$

Given the inputs $(x_1, x_2, ...., x_n)$ the model generates an output prediction of the movie rating. Where the inputs are first converted to Embeddings by mapping each input value to a vector as

$$e_i: x_i \mapsto \mathbf{x}_i \quad (2)$$

**Entity Embeddings** The model aims to map categorical variables in a function approximation problem into Euclidean spaces, which are the entity embeddings of the categorical variables. The mapping is learned by a neural network during the standard supervised training process (Guo and Berkhahn 2016). Entity Embeddings are advantageous in the setting of a movie recommendation system, as it performs well on datasets with lots of high cardinality features such as the User Id and Movie Id in the dataset, and more importantly, it reveals the intrinsic properties of the categorical variables which helps in the content-based clustering of movies and users. The embeddings are utilized to make recommendations personalized, by aggregating the movie embeddings of the movies a user rated highly, it is possible to extract the user's interests.

# Empirical Experiments and Results

The model is tested on a subset of the MovieLens dataset. The dataset contained 77167 entries, 5071 unique movies, and 671 unique users. The observations obtained are as follows: 1) A content-based approach performs well in the setting of a movie recommendation system. Table 2 shows the performance. 2) The embeddings learned was able to express the intrinsic properties of movies and users. 3) The system can be made personalized by aggregating embeddings of the user's top movies which captures the user's unique taste. Then searching the embedding space to return movies with features similar to the user's interest. An example is shown in Table 3, the model captures the user's interest in Romance and Comedy to recommend similar movies.

| MSE | RMSE | MAE |
|---|---|---|
| 0.387 | 0.622 | 0.45 |

Table 2: The evaluation metrics calculated by comparing the predicted output to the actual output are shown.

| User Id: 288 | |
|---|---|
| Top Rated Movies | Recommended Movies |
| The Fox and the Hound, 10 Things I Hate About You, 500 Days of Summer, 13 going on 30. An American Tail: Fievel Goes West | The Mirror Has Two Faces, Sleepless in Seattle, Syriana, The Notebook |

Table 3: Recommendations based on the user profile.

# Conclusion

This abstract proposes a recommender system using Entity Embeddings to be used in content-based environments like that of movie recommendation. This approach preserves the content of the item and can make recommendations personalized.